\documentstyle[12pt]{article}
\def\np{Nucl. Phys.}
\def\pl{Phys. Lett.}
\def\pr{Phys. Rev.}
\def\prl{Phys. Rev. Lett.}
\def\cmp{Comm. Math. Phys.}

\begin{document}
\baselineskip = 20pt

\def\footnotefont{\tenpoint}
\def\figures{\centerline{{\bf Figure
Captions}}\medskip\parindent=40pt%
\def\fig##1##2{\medskip\item{FIG.~##1.  }##2}}
\newwrite\ffile\global\newcount\figno \global\figno=1
\def\fig{fig.~\the\figno\nfig}
\def\nfig#1{\xdef#1{fig.~\the\figno}%
\writedef{#1\leftbracket fig.\noexpand~\the\figno}%
\ifnum\figno=1\immediate\openout\ffile=figs.tmp\fi\chardef\wfile=
\ffile%
\immediate\write\ffile{\noexpand\medskip\noexpand\item{Fig.\
\the\figno. }
\reflabeL{#1\hskip.55in}\pctsign}\global\advance\figno by1\findarg}

\parindent 25pt
\overfullrule=0pt
\tolerance=10000
\def\Re{\rm Re}
\def\Im{\rm Im}
\def\titlestyle#1{\par\begingroup \interlinepenalty=9999
     \fourteenpoint
   \noindent #1\par\endgroup }
\def\tr{{\rm tr}}
\def\Tr{{\rm Tr}}
\def\half{{\textstyle {1 \over 2}}}
\def\calt{{\cal T}}
\def\ie{{\it i.e.}}
\def\np{Nucl. Phys.}
\def\pl{Phys. Lett.}
\def\pr{Phys. Rev.}
\def\prl{Phys. Rev. Lett.}
\def\cmp{Comm. Math. Phys.}
\def\quart{{\textstyle {1 \over 4}}}
\def\RR{${\rm R}\otimes{\rm R}~$}
\def\NSNS{${\rm NS}\otimes{\rm NS}~$}

\baselineskip=14pt
\pagestyle{empty}
{\hfill DAMTP/96-12}
\vskip 0.4cm
\centerline{COMMENTS ON THREE-BRANES}
\vskip 1cm
 \centerline{ Michael B.  Green\footnote{M.B.Green@damtp.cam.ac.uk}
 and Michael Gutperle\footnote{M.Gutperle@damtp.cam.ac.uk}}
\vskip 0.3cm
\centerline{DAMTP, Silver Street,}
\centerline{ Cambridge CB3 9EW, UK.}
\vskip 1.4cm
\centerline{ABSTRACT}
\vskip 0.3cm
The Born--Infeld-like  effective  world-volume theory of a
single 3-brane is deduced from a manifestly space-time supersymmetric
description of the corresponding $D$-brane.
 This   is shown to be invariant under
$SL(2,R)$ transformations that act on the abelian gauge field  as well as the
bulk fields.  The
effective theory of two nearby parallel three-branes  involves  massive
world-volume supermultiplets which transform under $SL(2,Z)$ into the dyonic
solitons of  four-dimensional $N=4$ spontaneously broken $SU(2)$  Yang--Mills
theory.

\vfill\eject
\pagestyle{plain}
\setcounter{page}{1}

The $p$-brane solitonic solutions of string theory  play an important r\^ole in
determining the systematics of the duality relationships between string
theories.       These $p$-branes  carry charges associated with $p+1$-form
potentials in the Neveu--Schwarz-Neveu--Schwarz (\NSNS) sector and the
 Ramond--Ramond \RR sector.  According to the $D$-brane description
\cite{polchina} the \RR sector
 $p$-branes  are described by configurations of open superstrings
with end-points tethered on a $(p+1)$-dimensional hypersurface embedded in ten
dimensions.  The surface may, in principle, be curved but it is simplest in the
first instance to consider flat $p$-branes.    The coordinates of the
open-string end-points are restricted to the plane $X^i = y^i$, where the
directions labelled $i$ are transverse to the brane world-volume ($i =p+1,
\cdots, 9$).  The dynamics of the brane is  therefore prescribed by the open
superstring theory with Neumann boundary conditions in the directions labelled
by $\alpha =0,\dots, p$ and Dirichlet boundary conditions in the transverse
directions \cite{daipolchinski,wittenb}.
The usual kind of  \lq effective' world-volume field theory emerges only as an
approximation to this \lq underlying'  open superstring theory.    A
characteristic feature of the world-volume theory of a \RR $p$-brane is the
ocurrence of a world-volume vector field \cite{callanl,duffc}.  In the
$D$-brane description this arises as a ground state of the open-string
theory.   The fact
that  such $D$-branes are described by {\it open} superstring theory means that
they preserve half the supersymmetry of the fundamental  type II string theory,
which means that they are  BPS solitons of the type II theories.
  The supersymmetries of the type II theory that are broken by the open-string
truncation are associated with the Goldstone fermions living in the
world-volume \cite{polchinhughes}.

\vskip 0.2cm
\noindent{\it Source equations and effective actions}
\vskip 0.1cm

The  effective action for superstring theory  in the presence of a solitonic
\RR $p$-brane can, in principle,  be written in the form
\begin{equation}\label{stot}
S = S_{bulk} + S^{(p)}_{source},
\end{equation}
where the bulk action is a ten-dimensional integral while the source is
restricted to $p+1$ dimensions.   The equations of motion that follow from this
action arise in  the underlying string theory from the consistency conditions
that ensure conformal invariance of the world-sheet theory.  In the case of the
type IIB theory there is no obvious covariant bulk action (due to the presence
of a five-form field strength that satisfies a self-duality condition) but the
source-free equations   of motion are known \cite{schwarzx,howea}.   The bulk
terms
arise from the usual closed-string sector of the theory but the source terms
come from world-sheets with boundaries on which there can be a condensate of
the open-string fields.   These actions, which  are of the Born--Infeld type,
reproduce the  equations of motion that follow from the consistency conditions
for string theory in the background that includes the open-string  boundary
condensate, $F=dA$, of the electromagnetic field where $A$ is  the massless
open-string abelian vector potential.  Such conditions
on  open string theory with Neumann conditions in all directions were
originally derived  \cite{fradkina,callana,polchinq} in the
ten-dimensional type I theory assuming that $F_{\alpha\beta}$  is constant.
Imposing Dirichlet conditions in $9-p$ directions defines a  $D$-brane boundary
state.

An effective action of the form (\ref{stot}) is  obtained in the case of the
\RR one-brane (the $D$-string) by considering the consistency conditions for
type IIB string theory in the presence of  world-sheet boundaries   with
Neumann conditions in
two  directions  and fixed  Dirichlet  boundary conditions in the remaining
eight  directions.    The form of  $S^{(1)}_{source}$ is determined in the case
that  $F$ is constant to be a version  of the Born--Infeld theory coupled  to
the metric (G),
the background  antisymmetric potentials ($B^N$ and  $B^R$) and  scalars
($\phi$ and $\chi$)   in the \NSNS and \RR sectors.     It was shown in
\cite{schmidhubera,alwisa} that this is the appropriate world-sheet action to
describe a   type IIB string carrying charges in the
\NSNS and \RR sectors.  Furthermore,   the action of $SL(2,Z)$  on the
background closed-string fields induces the transformation among the infinite
set of dyonic  type IIB strings that arise as solitons in the usual
field-theoretic construction \cite{schwarzx}.

In this paper we will use the constraints of space-time supersymmetry
(formulated in a light-cone gauge) to determine  the  effective  world-volume
theory of  the three-brane which is a four-dimensional supersymmetric theory
with $N=4$ global supersymmetry. The  bosonic part of the three-brane theory
will be described and the $SL(2,R)$ invariance of its equations of motion
demonstrated.   Any vacuum  configuration of  $n$ parallel $p$-branes
describes
the moduli space of
supersymmetric  $U(n)$ Yang--Mills theory \cite{wittenb}  dimensionally
reduced  from ten
dimensions to $p+1$ dimensions where the   $U(1)$ symmetry of a particular
brane can be viewed as a subgroup of $U(n)$.   The system of two closely
separated parallel three-branes has excited states in which the branes are
joined by a
string which may be either a fundamental type IIB string or any one of the
dyonic solitonic type IIB strings \cite{stromingerb}.  We will discuss how
$SL(2,Z)$
acts  to reveal the usual spectrum of solitonic dyonic states in the
four-dimensional world volume.

\vskip 0.2cm
\noindent{\it  Light-cone boundary states  and D-branes.}
\vskip 0.1cm

A  closed-string state  in the \RR
or \NSNS sector  coupling to a world-sheet with a boundary  can be represented
as a semi-infinite
cylinder that describes the evolution of a closed string state
$|\Phi\rangle$ from $\tau =- \infty$ to the boundary at $\tau =0$,
\begin{equation}\label{diskcoup}
\langle \Phi \rangle_{Disk} = \langle \Phi|B, \eta\rangle,
\end{equation}
where $|B,\eta\rangle$ denotes the boundary state and  $\eta = \pm 1$  labels
whether the state is BPS or anti-BPS.       In the formalism with manifest
world-sheet
supersymmetry there are separate boundary states in each sector, $|B,
\eta\rangle_N$ and $|B, \eta\rangle_R$,  which  both satisfy the boundary
conditions,
\begin{equation}\label{boses}
 (\partial X^i -\bar \partial X^i)|B,\eta\rangle =0, \qquad  (\partial X^\alpha
+ \bar\partial X^\alpha)|B,\eta\rangle=0
\end{equation}
that impose Neumann conditions  on $p+1$ directions and Dirichlet conditions on
the rest (a state  $|B,\eta\rangle$ with no subscript is in either of the two
possible sectors).   Implicit in (\ref{boses}) is the fact that $X^i(\sigma)|B,
\eta\rangle = y^i |B,\eta\rangle$, where $y^i$ is the constant transverse
position of the brane.    In order to ensure that half the world-sheet
supersymmetry
is preserved the  world-sheet fermions must satisfy the boundary conditions
\begin{equation}\label{ferm}
 (\psi^i + i\eta \tilde \psi^i)|B,\eta\rangle=0, \qquad  (\psi^\alpha- i\eta
\tilde \psi^\alpha )|B,\eta\rangle=0.
\end{equation}
It is easy to see that these conditions ensure that an infinite sequence of
local supersymmetry gauge conditions are preserved   \cite{greeng,polchina},
\begin{equation}\label{supergauge}
(F  + i\eta \tilde F) |B,\eta\rangle =0,
\end{equation}
where $F=  \psi\cdot  \partial X$, $\tilde F =\bar \psi \cdot  \bar
\partial X$.

Space-time supersymmetry, which is not manifest in the  above approach, relates
the two kinds of boundary state, $|B\rangle_N$ and $|B\rangle_R$.
However, it is also possible to formulate the problem in a  light-cone gauge
 in which  space-time supersymmetry is manifest \cite{greeng,  guta}.   In this
gauge $\tau= X^+= (X^0+X^9)/\sqrt 2$ so that the boundary state is at a fixed
time, i.e,  $X^+$ is
required to satisfy a Dirichlet condition.  Furthermore, $X^-= (X^0 -
X^9)/\sqrt 2$  is
determined in terms of the transverse $X^I$ coordinates ($I=1,2,\cdots,8$) and
their fermionic partners and also satisfies a Dirichlet condition  (whether
$X^I$ are Neumann or Dirichlet
coordinates).   This means that there are at least two Dirichlet directions and
that one of these is time-like.  This kinematics describes a \lq
(p+1)-instanton' rather than a $D$-brane, where time is one of the Neumann
directions.  It is related to the $D$-brane by a double Wick rotation.
The  coordinates transverse to the $\pm$ directions  satisfy the
boundary conditions
\begin{equation}\label{bosboun}
(\partial X^I + M_{IJ} \tilde\partial X^J) |B,\eta\rangle =0,\end{equation}
where $M_{IJ}$ is an element of  $SO(8)$.  The Neumann directions will be
chosen to be $\alpha = I =1, \cdots, p+1$ while the Dirichlet directions will
be chosen to be $i = I = p+2, \cdots ,8$.   In the absence of a boundary
condensate this matrix can be written in block diagonal form,
\begin{equation}\label{mijdef}
M_{IJ} = \pmatrix { I_{p+1}& 0 \cr
      0 & -I_{7-p}  \cr}
\end{equation}
(where $I_q$ indicates the $(q\times q)$-dimensional unit matrix).  In the
presence of a boundary condensate of the open-string vector
potential $M_{IJ}$  is a more general $SO(8)$ matrix and can be written as,
\begin{equation}\label{so8vec}
M_{IJ} = -\exp\left\{\Omega_{KL}  \Sigma^{KL}_{IJ} \right\},
\end{equation}
where   $\Sigma^{KL}_{IJ} =  (\delta^K_{\ I}\delta^L_{\ J} - \delta^L_{\
I}\delta^K_{\ J}) $ are generators of $SO(8)$ transformations in the vector
representation.  The parameters  $\Omega_{IJ}$ depend on the open-string
boundary condensate and can be written in block off-diagonal form in a
particular basis as,
\begin{equation}\label{omdef}
\Omega_{\alpha\beta} = \sum_{ m=0}^{(p-1)/2}  c_m \Sigma^{2m+1\
2m+2}_{\alpha\beta},
\end{equation}
and $\Omega_{ij}= \Omega_{\alpha i} =0$   if the brane is static.  In the
absence of a condensate $c_m = \pi$.

The conditions on the fermionic coordinates will be derived by requiring the
boundary state to be annihilated by a linear combination of the space-time
supercharges,
\begin{equation}\label{susyi}
Q_\eta^{+a}  \equiv (Q^a + i\eta M^a_b \tilde Q^b)|B,\eta\rangle =0, \qquad
Q_\eta^{+ \dot a} \equiv (Q^{\dot a} + i\eta
M^{\dot a}_{\dot b} \tilde Q^{\dot b})|B,\eta\rangle =0,
\end{equation}
where $Q^a$ and $Q^{\dot a}$ are two inequivalent $SO(8)$ left-moving
supercharges and $\tilde Q^a$, $\tilde Q^{\dot a}$ are the right-moving ones.
These conditions enforce the requirement that  the boundary state preserves one
half of the
space-time supersymetries, generalizing one of the arguments in  \cite{greeng}
to the context of
the $D$-brane.   This means that the $D$-brane is a BPS saturated state.

It is easy to deduce from (\ref{susyi}) that the modes of the   $SO(8)$
fermionic world-sheet fields,  $S^a$ and $S^{\dot a}$,    must satisfy the
boundary conditions,
\begin{equation}\label{modsusy1}
(S_n^a+i\eta M_{ab}\tilde{S}^b_{-n})| B, \eta \rangle =  0, \qquad
(S_n^{\dot{a}}+i \eta M_{\dot{a}\dot{b}}\tilde{S}^{\dot{b}}_{-n})| B,
\eta\rangle = 0.
\end{equation}

The   bispinor $SO(8)$  matrix $M_{ab}$ is determined by consistency with
the superalgebra to satisfy the condition
\begin{equation}\label{mconst}
M_{ab}M_{cb} = \delta_{ac},
\end{equation}
and $M_{\dot a\dot b}$ satisfies,
\begin{equation}\label{mdotsat}
\gamma^I_{a\dot a}   - M_{\dot a\dot b} M_{ba}  M_{JI} \gamma^J_{b\dot b} =0.
\end{equation}
These conditions imply that $M_{ab}$ and $M_{\dot a\dot b}$ describe the same
$SO(8)$ rotations  as $M_{IJ}$ but acting on the  spinors rather than on
vectors.
In the absence of a condensate these equations are solved by
\begin{equation}\label{mdef}
M_{ab}  = \left(\gamma^1\gamma^2 \dots \gamma^{p+1}\right)_{ab},  \qquad
M_{\dot a\dot b}  = \left(\gamma^1\gamma^2 \dots \gamma^{p+1}\right)_{\dot
a\dot b}.
\end{equation}
 More generally, in the presence of a condensate,
\begin{equation}\label{so8spin}
M_{ab} = e^{\half \Omega_{IJ} \gamma^{IJ}_{ab}},
\end{equation}
with the same $SO(8)$ rotation $\Omega_{IJ}$ as before.

The boundary state that satisfies these conditions is given by
\begin{equation}\label{boundarystate2}
| B\rangle= \exp\sum_{n>0} \left( {1\over n} M_{IJ}
\alpha^I_{-n}\tilde{\alpha}^J_{-n}   + i M_{ab}S^a_{-n}\tilde{S}^a_{-n}\right)
 |B_0\rangle
\end{equation}
where the zero-mode factor is
\begin{equation}\label{boundarystate1}
| B_0\rangle =C\left( M_{IJ}| I\rangle\tilde{|
J\rangle}+iM_{\dot{a}\dot{b}}| \dot{a}\rangle\tilde{|
\dot{b}\rangle}\right).
\end{equation}
The overall normalization constant $C$ of the boundary state can easily be
determined by relating the cylinder diagram to an annulus (as in
\cite{polchinq,aboula}) which is here interpreted
as the free energy of a gas of open strings and is given as a trace over
physical open-string states.   This boundary state defines source terms for all
the massless and massive closed-string fields since it  describes the relative
couplings of the boundary to the closed-string  states.  The source terms for
the massless fields in the effective action can be obtained from  the zero-mode
factor  (\ref{boundarystate1}).  The results agree with calculations in the
covariant boundary state formalism in the  purely Neumann case considered in
\cite{callanb} and the $D$-string considered in \cite{lia,alwisa}.

The non-linearly realized supersymmetries are those that are generated by the
supercharges that are not annihilated by the boundary state, $Q_\eta^{-a}
\equiv (Q^a - i\eta
M^a_b \tilde Q^b)$ and $Q_\eta^{-\dot a} \equiv (Q^{\dot a}  - i\eta M^{\dot
a}_{\dot b} \tilde Q^{\dot  b})$.   The $Q^-$ supercharges are zero-momentum
fermion emission vertices so that $Q^{-a}|B\rangle$ and $Q^{-\dot a}| B
\rangle$ are equivalent to  zero-momentum fermion insertions and a soft
Goldstino theorem can be demonstrated \cite{guta} as required for a
non-linearly realized  supersymmetry.

\vskip 0.2cm
\noindent{\it  The three-brane source equations}
\vskip 0.1cm

In  the example of the  three-brane (or four-instanton) there are  four
transverse Dirichlet directions and four Neumann (recall that in this
discussion the $\pm$ light-cone directions are identified with two directions
transverse to the $(p+1)$-dimensional world-volume).

 In the absence of a boundary condensate  it is easy to see that the boundary
state (\ref{boundarystate1}) does not couple to the dilaton.  This follows
simply from the fact that $\Tr M_{IJ} =0$   (with $M_{IJ}$ defined in
(\ref{mijdef})).   This agrees with the ansatz that defines the three-brane
soliton of the supergravity in \cite{duffc}  where the only non-trivial fields
are the graviton and the fourth-rank antisymmetric potential $A^{(4)}$.
Furthermore, since $M_{ab} = \half(\gamma^{1234} + \gamma^{5678})$ the boundary
state couples equally to $A^{(4)}$ and $\hat * A^{(4)}$ (where $\hat *$ denotes
the Poincar\'e dual in the eight-dimensional space transverse to the $\pm$
directions) and therefore it respects the self-duality of
$F^{(5)}$.

More generally, there will be a boundary condensate of the vector potential
with field strength $F_{\alpha\beta}$, which will here be assumed to be
constant.     In this case the $SO(8)$
 matrix, $M_{IJ}$, may be written in $4\times 4$  block diagonal form in a
suitable basis  \cite{lia,guta},
\begin{equation}\label{mathree}
M_{IJ} = \pmatrix{{(1-F) \over  (1+F)}   & 0 \cr
           0 & -1_4\cr}
\end{equation}
where $F\equiv F_{\alpha\beta}$ is the constant $4\times 4$  field strength in
the world-volume of the brane.
An orthogonal transformation brings this to the form,
\begin{equation}\label{fmatr}
F_{\alpha\beta} = \pmatrix{ 0&f_1 & 0 & 0 \cr
        -f_1 & 0& 0& 0 \cr
          0&0&0&f_2 \cr
        0& 0 & -f_2 &0 \cr}
\end{equation}
so that,
\begin{equation}\label{blocdiag}
 {1-F\over 1+F}   = \pmatrix{M(f_1) & 0 \cr
       0 & M(f_2) \cr}
\end{equation}
where
\begin{equation}\label{matrixM}
M(f) = \frac{1}{1+f^2}
\pmatrix{ 1-f^2 &-2f \cr
2f&1-f^2\cr}
\end{equation}
The  matrix $M_{IJ}$  can be  expressed in the form (\ref{so8vec}) with $c_n =
\pi + \alpha_n$, where $\cos \alpha_n = (1-f_n^2)/(1+f_n^2)$ (with $n=1,2$).
Since $M_{IJ}$ has both symmetric and antisymmetric parts it is a source for
$B^N$ as well as the metric and dilaton.

Likewise the matrix $M_{ab}$ is modified by the boundary condensate and can be
written in the same basis as
\begin{equation}\label{matab}
M_{ab}={1 \over  \sqrt{(1+f_1^2)(1+f_2^2)}  }
(f_1f_2\delta_{ab}+f_2\gamma_{ab}^{12}+f_1\gamma_{ab}^{34}+\gamma^{1234}_{ab})
\end{equation}
The normalization constant is determined by considering the equivalence of the
annulus and cylinder diagrams and is   $C=
\sqrt{(1+f_1^2)(1+f_2^2)}= \sqrt{\det(1+F)}$.

\vskip 0.2cm
\noindent{\it  The effective three-brane action and $SL(2,R)$ symmetry.}
\vskip 0.1cm

It is straightforward to verify that the covariant (euclidean)  effective
action that reproduces these source terms has the  form,
\begin{equation}\label{effact}
S^{(3)}_{source} = \int d^4x \left(e^{-\phi} \sqrt{\det (G+{\cal F})}  +
{i\over
4} \chi {\cal F} * {\cal F} + {i\over 2}  B^R * {\cal F} +{i\over 24}
\epsilon^{\alpha\beta\gamma\delta}  A^{(4)}_{\alpha\beta\gamma\delta}  \right),
\end{equation}
where ${\cal F} = F -B^N$ and $*F_{\gamma\delta} =\half
\epsilon^{\alpha\beta}_{\ \ \gamma\delta}
F_{\alpha\beta}$.
The sources generated by the boundary state (\ref{boundarystate1}) match up
with the expansion  of (\ref{effact}) in  small fluctuations of the bulk
closed-string  fields to linearized order, keeping all orders in $F$.  The \RR
sector terms, proportional to $\chi$, $B^R$ and $A^{(4)}$, match in an obvious
manner. The terms in the \NSNS sector are proportional to $\phi$ and
$h_{\alpha\beta} = \eta_{\alpha\beta} - G_{\alpha\beta} +   B_{\alpha\beta}^N$.
It is important to remember that in the light-cone gauge $\phi = h^I_{\ I} /4 =
 (h^i_{\ i} + h^\alpha_{\ \alpha})/4$.   The expansion of the
determinant factor in (\ref{effact}) gives
\begin{equation}\label{detfact}
-\phi \sqrt{\det(1+F)} + \half \sqrt{\det(1+F)} h^{\alpha\beta}
(1+F)^{-1}_{\alpha\beta} =
{1\over 4} \sqrt{\det(1+F)}  \left(- h_{ii} + \left({1-F\over 1+F}\right)
_{\alpha\beta} h^{\alpha\beta}\right),
\end{equation}
which is precisely the same as the source obtained from the boundary state
(\ref{boundarystate1}).

   The action (\ref{effact}) encodes the information about the
equations of motion that follow from the string consistency conditions.
However, it is only these equations that we shall make use of in what follows
since it is well known that the presence of the self-dual five-form field
strength means that the bulk part of the action (\ref{stot}) cannot be written
in a covariant manner.   This action is presented in the string frame in
(\ref{effact}).  In order to discuss the duality
properties of the theory we shall transform this to the Einstein frame by
transforming $G\to G_E =  Ge^{-\phi/2}$.  This replaces  (\ref{effact}) with
\begin{equation}\label{newact}
S^{(3)}_{source} = \int d^4x  \left( \sqrt{\det (G_{E}+e^{-\phi/2} {\cal F})}
+{i\over 4}  \chi {\cal F}  * {\cal F}  + {i \over 2}  B^R *  {\cal F} +{i\over
24} \epsilon^{\alpha\beta\gamma\delta}  A^{(4)}_{\alpha\beta\gamma\delta}
\right),
\end{equation}
with the corresponding change of metric in the bulk part of the total action.

When continued to lorentzian signature the  action (\ref{newact}) is remarkably
similar to an action studied in   \cite{gibbonsb} which was
not related to consideration of $D$-branes.  There, the bulk term was supposed
to be  a four-dimensional  $SL(2,R)$-invariant theory while the Born--Infeld
part, $S_{source}^{(3)}$,  was precisely of the form (\ref{newact})  with the
fields $B^N$ and $B^R$ set equal to zero.  The pseudoscalar was referred to as
the axion in \cite{gibbonsb}  while in (\ref{newact}) it is identified with the
\RR scalar, $\chi$.

In the absence of the antisymmetric tensor fields the   $SL(2,R)$
transformations  act on the fields in the action  of  \cite{gibbonsb} as
follows,
\begin{eqnarray}
\lambda & \to& \frac{p\lambda+q}{r\lambda+s}\;\\
F_{\mu\nu}&\to& sF_{\mu\nu}+r*G_{\mu\nu}\label{Ftrafo} ,\\
G_{\mu\nu}&\to& pG_{\mu\nu}-q*F_{\mu\nu}\label{Gtrafo}\\
\Lambda&\equiv &\left(\begin{array}{cc}
p&q\\
r&s\end{array}\right)\in SL(2,R),
\end{eqnarray}
where $\lambda =\chi+ie^{-\phi}$,  $G$ is defined by
\begin{equation}\label{gdef}
G_{\mu\nu}=-2\frac{\delta S^{(3)}_{source}}{\delta F^{\mu\nu}},
\end{equation}
(and  $** =-1$  with Lorentzian signature).

The invariance of the equations of motion can be seen by considering
infinitesimal $SL(2,R)$ variations with $p=1+\alpha$,  $q= \beta$, $r=\gamma$
and $s=1-\alpha$,
\begin{eqnarray}\label{varfield}
 \delta\chi = 2\alpha\chi+\beta-\gamma(\chi^2-e^{-2\phi}) , &\qquad \delta\phi
=  2(\chi\gamma-\alpha)\\
 \delta F_{\mu\nu} = \gamma *G_{\mu\nu}-\alpha F_{\mu\nu}, &\qquad \delta
G_{\mu\nu} = \alpha G_{\mu\nu}-\beta *F_{\mu\nu}
\end{eqnarray}

The steps outlined in \cite{gibbonsb} that demonstrate the $SL(2,R)$ invariance
of
the equations of motion of the theory may now  be generalized to include the
extra background fields  $B^N$ and $B^R$ in (\ref{newact}) which transform as,
\begin{equation}  \label{duality}
\left(\begin{array}{c}
B^N\\
B^{R}
\end{array}\right)\to(\Lambda^{-1})^T\left(\begin{array}{c}
B^N\\
B^{R}
\end{array}\right)= \left(\begin{array}{cc}
s&-r\\
-q&p
\end{array}\right)\left(\begin{array}{c}
B^N\\
B^{R}
\end{array}\right)
\end{equation}
and the fourth-rank potential, $A^{(4)}$,  which is invariant.
The infinitesimal form of these transformations is,
 \begin{equation}\label{brbns}
 \delta B^N  =  -\gamma B^R_{\mu\nu}-\alpha B^N_{\mu\nu} ,  \qquad \delta
B^{R}_{\mu\nu} = \alpha B^{R}_{\mu\nu}-\beta B^N_{\mu\nu}
\end{equation}

The
$SL(2,R)$ symmetry of the equations of motion is dependent on the
invariance of $G$ which was defined by (\ref{gdef}). Its variation is given by
\begin{equation}
\delta G_{\mu\nu}= \left( \frac{\partial G_{\mu\nu}}{\partial
  F_{\rho\sigma}}\delta F_{\rho\sigma}+\frac{\partial G_{\mu\nu}}{\partial
  B^N_{\rho\sigma}}\delta
B^N_{\rho\sigma}+\frac{\partial G_{\mu\nu}}{\partial\chi}\delta
\chi+\frac{\partial G_{\mu\nu}}{\partial\phi}\delta \phi\right)+\delta
*B^R_{\mu\nu}\label{vari}
\end{equation}

There are  three independent  terms in the transformations with coefficients
$\beta,\gamma,\alpha$.
The $\beta$ term  in (\ref{vari}) gives
\begin{equation}\label{beta}
-*F_{\mu\nu}=  \frac{\partial
  G_{\mu\nu}}{\partial\chi}-*B^N_{\mu\nu},
\end{equation}
and since     $  \partial G_{\mu\nu} / \partial\chi = - *{\cal F}_{\mu\nu} =
-*(F-B^N)_{\mu\nu}$,   this equation is identically satisfied.

The $\gamma$ term in  (\ref{vari}) gives
\begin{equation}
0= \left(\frac{\partial G_{\mu\nu}}{\partial
  F_{\rho\sigma}}(*G_{\rho\sigma}   +
B^{R}_{\rho\sigma})-(\chi^2-e^{-2\phi})\frac{\partial
G_{\mu\nu}}{\partial\chi}+2\chi\frac{\partial G_{\mu\nu}}{\partial\phi}\right),
\end{equation}
and the $\alpha$ term gives
\begin{equation}
G_{\mu\nu}=  \left(\frac{\partial G_{\mu\nu}}{\partial
  F_{\rho\sigma}}(-F_{\rho\sigma}+B^N_{\rho\sigma})+2\chi\frac{\partial
  G_{\mu\nu}}{\partial\chi}+2\chi\frac{\partial
  G_{\mu\nu}}{\partial\phi}\right) +  *B^{R}_{\mu\nu}.
\end{equation}

These variations reduce to the ones given  in \cite{gibbonsb}  but
with $F$ replaced by ${\cal F}$.  Therefore, the arguments in \cite{gibbonsb}
generalize to demonstrate the $SL(2,R)$
invariance of the equations in the presence of $B^N$ and $B^R$ which is broken
to $SL(2,Z)$ in the quantum theory.   Although the action is not
uniquely determined by these arguments the requirements of supersymmetry
constrain it considerably -- not only must the theory possess manifest
$N=4$ supersymmetry but there must also be four non-linearly realized
supersymmetries.

The consistency of  $p$-branes  ending in $(p+2)$-branes was discussed in
general terms in  \cite{stromingerb,townsenda}.  In this case we
consider a dyonic string
soliton carrying  charges $(Q^N,Q^R)$ associated with the \NSNS sector and the
\RR sector  terminating on a three-brane.   An  eight-sphere $S^8$  intersects
the string at
a point and the three-brane on a two-sphere  $S^2$  surrounding the end-point
of the
string.  The equations of motion for the antisymmetric tensor    field
strengths in the presence of the string and the three brane
are given by (in form notation)
\begin{eqnarray}
d\tilde{*}H^{R}&=&Q^{R}\delta^8(x)+\frac{\delta S^{(3)}}{\delta B^R}\wedge
\delta^6(x)\\
d\tilde{*}H^N&=&Q^N\delta^8(x)+\frac{\delta S^{(3)}}{\delta B^N}\wedge
\delta^6(x)
\end{eqnarray}
where the $\delta^8$ and $\delta^6$ are the 8 and 6 form delta
functions in the space transverse to the string and the three-brane
respectively (and $\tilde{*}$ denotes duality with respect to the
ten-dimensional space-time). Using the form of the three-brane action and
integrating
over the $S^8$ gives
\begin{equation}\label{charges}
0 =  Q^R + \int_{S^2}{\cal F} , \qquad 0=
Q^N +  \int_{S^2} *G\label{charge2}
\end{equation}
where $G$ is defined by (\ref{gdef}). This shows that the string end-point  in
the four-dimensional world-volume  is a dyon carrying
electric and magnetic charges that are equal to  $Q^N$ and $Q^R$,
respectively.  A   $SL(2,Z)$ duality transformation  that acts on the
antisymmetric
tensor charges by
\begin{equation}
\left(\begin{array}{c}
Q^N\\
Q^{R}
\end{array}\right)\to\Lambda\left(\begin{array}{c}
Q^N\\
Q^{R}\end{array}\right)= \left(\begin{array}{cc}
p&q\\
r&s
\end{array}\right)\left(\begin{array}{c}
Q^N\\
Q^{R}
\end{array}\right)
\end{equation}
also acts on the worldvolume vector potential  as a  duality transformation
of  the type considered in \cite{sena}.  This transforms the
fundamental electric
charges into the infinite set of dyonic charges.

It is easy to see that the
combined system of a string terminating in a three-brane breaks half of the
space-time supersymmetry of the single three-brane.  This means that it breaks
three quarters of the original 32-component type IIB supersymmetry
leaving eight unbroken supercharges, which is
consistent with the presence of dyonic solitons in four-dimensional $N=4$ super
Yang--Mills theory.

\vskip 0.2cm
\noindent{\it   Parallel three-branes  and dyons.}
\vskip 0.1cm

It  was emphasized in   \cite{wittenb} that  the configuration space of  $n$
parallel $p$-branes describes the moduli space of ten-dimensional
supersymmetric $U(n)$ Yang--Mills theory dimensionally reduced to $p+1$
dimensions.  If
the branes are not coincident the $U(n)$ symmetry breaks to a subgroup and if
none of them are coincident it breaks to $U(1)^n$ -- one  $U(1)$ factor for
each brane.  The broken generators of the group
correspond to massive gauge potentials that are the ground states of strings
that stretch from one brane to
another, $A_\mu^{rs}$ (where $r,s$ label the branes on which the endpoints are
fixed and $\mu = 0,\cdots,9$)  while the unbroken $U(1)$'s are associated with
open strings with both
end-points terminating on the same brane that have massless ground states that
are the  potentials $A_\mu^{rr}$.

Here we will consider two parallel three-branes  with world-volume coordinates
in directions $\alpha =0,\cdots,3$ and transverse coordinates in directions
$i=4,\cdots,9$.   The   $U(2) =
SU(2) \times U(1)_D$ symmetry  is broken to $U(1)_A \times U(1)_D$ when the
branes are separated. The diagonal  abelian group, $U(1)_D$, is associated with
an overall phase and
has gauge connection $A_\mu^+ =\half( A_\mu^{11} + A_\mu^{22})$ while $U(1)_A$
transforms the relative phase and its connection is $A_\mu^- = \half(A^{11}_\mu
- A^{22}_\mu)$.  The $A^{rr}_\alpha$ ($\alpha=0,\cdots, 3$) are the massless
world-volume potentials in each of the two branes while $\phi^{r}_i = A^{rr}_i$
($i=4,\cdots,9$) are the massless scalar fields  that describe the transverse
positions of the two branes. Hence $\phi^+_i$ is the center of mass position
and $\phi^-_i$ is the relative position of the branes.

The supersymmetric ground state of this system is parameterized by the
separation between the branes  in a particular transverse direction, which
will be  taken to be $i=9$.  We will be interested in closely spaced
branes with separation  $R<< (\alpha')^{1/2}$.   In other words $|\langle
\phi^-_9\rangle | =
R$.  This non-zero expectation value of  a scalar field is a symptom of the
fact that
the $U(2)$ symmetry is broken spontaneously. The fact that this scalar field is
massless is characteristic of a BPS saturated system which has flat
directions in the potential.   None of these massless
fields is charged under the $U(1)$'s.

The system also possesses excited states in which the two branes are joined by
a  stretched fundamental string with endpoints moving in the branes.  The
string world-sheet lies
in the $0-9$ plane. A fundamental string stretching between the three-branes
has
ground-state vector potentials, $A_I^{\pm}(x^\alpha) = \half(A_I^{21}(x^\alpha)
\pm A_I^{12}(x^\alpha))$, where $I=1,\cdots,8$
are the eight directions transverse to the open-string world-sheet.   Five of
these directions are transverse to both the world-volume of the three-brane and
the world-sheet of the stretched  string  and the other three directions lie in
the  three-brane.  The latter three components are interpreted as the physical
components of a massive four-dimensional world-volume vector $W^\pm (x^\alpha)$
  -- a $W$-boson --
and the other
components, $\rho^{\pm A}(x^\alpha)$  ($A=1,\cdots, 5$) are the five scalar
fields needed to fill out the bosonic components  of a massive short
world-volume $N=4$  supermultiplet.  There  are also eight
massive ground-state  fermionic fields.    Therefore the world-volume theory
contains two charged massive supermultiplets, in addition to the massless
fields considered above.  These  massive multiplets carry charges $\pm 1$ under
$U(1)_A$.
The mass of these states depends on the separation of
the branes,
\begin{equation}\label{separmass}
M =(\alpha')^{-1}  R.
\end{equation}
But this is precisely the content of the usual
$N=4$  four dimensional supersymmetric $U(2)$ Yang--Mills theory with
spontaneous symmetry
breaking, where the scale of the breaking is determined by the vacuum value of
a massless scalar field, $\phi^-_9$.

Instead of  the fundamental string joining the branes   a
solitonic string  can join them.   This can be any of the infinite number of
\lq
dyonic'  type IIB strings  \cite{schwarzb} with string tensions depending on
the
values of the charges $m$ and $n$  that couple to the antisymmetric tensors,
$B_{09}^N$ and $B_{09}^R$, respectively,
\begin{equation}\label{tensio}
T_{m,n} =   {1\over \alpha'}  \left((m-n\chi)^2 + {n^2 \over g^2}\right)^{1/2}.
\end{equation}
Since the tension of the solitonic string
is very large in pertubation theory ($T_{m,n} \sim 1/g$) the string length is
the minimum possible
length, which is $R$ and so the positions of the end-points are at $x^{1\alpha}
=
x^{2\alpha} = x^\alpha$.   Using Gauss' law in the form (\ref{charges})  at
either end-point of the stretched string shows that $e_1=-e_2$, where $e_r$ is
the charge that couples to the $U(1)$ potential $A^{rr}$.  Therefore the
diagonal  $U(1)_D$ charge  ($e_1$ + $e_2$)  vanishes but the relative $U(1)_A$
charge ($e_1 - e_2$) is non-zero.  This means that the
effective four-volume theory contains massive solitonic dyons that arise from
the composite two-brane system. The action of $SL(2,Z)$ transforms the dyons
among themselves, at the same time transforming
the strings that connect the branes.

This description of the dyonic spectrum of the $N=4$ supersymmetric
Born--Infeld theory  is
quite different from that of the version of $N=4$ supersymmetric
Yang--Mills theory  obtained from toroidal compactification of the
ten-dimensional heterotic theory \cite{sena}.  There the $W$ boson mass has a
scale $1/R$, where $R$ is the radius of a compactified dimension, and the
theory does not have an obvious non-linearly realized supersymmetry.
The  representation of spontaneous symmetry
breaking in terms of  two copies of the  four-dimensional world-volume is
somewhat  reminiscent of  \cite{connesa}.

Effective world-volume theories obtained from $D$-branes (such as the
three-brane  considered in this paper) are determined by
the  underlying open superstring theory.  Intriguingly, the quantum corrections
to the world-volume theory are given by open-string loop diagrams which
unavoidably induce the closed-string gravitational sector with states that
propagate  in the ten-dimensional embedding space.  Thus, the  distinction
between the effective world-volume and the embedding space should disappear in
the full quantum theory.

\vskip 0.3cm
{\it Note Added:}
During the preparation of this manuscript  a paper appeared in the hep-th
archive that has some overlap with the material presented here
\cite{tseytlinx}.
\vskip 0.2cm
{\it Acknowledment:}
M.Gutperle is grateful for financial support from the EPSRC and a Pannett
research studentship from Churchill College, Cambridge.

\end{document}